\documentclass[aps,prd,preprint,superscriptaddress,amsmath,amssymb,showpacs]{revtex4-1}
\usepackage{dcolumn}
\usepackage{graphicx}
\usepackage{float}
\usepackage{physics}
\usepackage[colorlinks=true,allcolors=blue]{hyperref}

\begin{document}

\title{$R^2$ corrections to holographic heavy quarkonium dissociation}

\author{Zhou-Run Zhu }
\email{zhuzhourun@zknu.edu.cn}
\affiliation{School of Physics and Telecommunications Engineering, Zhoukou Normal University, Zhoukou 466001, China}

\author{Manman Sun }
\email{sunmm@zknu.edu.cn}
\affiliation{School of Physics and Telecommunications Engineering, Zhoukou Normal University, Zhoukou 466001, China}

\author{Rui Zhou }
\email{ruychou@zknu.edu.cn}
\affiliation{School of Physics and Telecommunications Engineering, Zhoukou Normal University, Zhoukou 466001, China}

\author{Jinzhong Han}
\email{hanjinzhong@zknu.edu.cn}
\affiliation{School of Physics and Telecommunications Engineering, Zhoukou Normal University, Zhoukou 466001, China}

\begin{abstract}
In this paper, we investigate the $R^2$ corrections to the dissociation of heavy quarkonium in the Gauss-Bonnet gravitational background. We analyze the impact of Gauss-Bonnet parameter $\lambda_{GB}$ on the spectral function of charmonium and bottomonium, and examine how $\lambda_{GB}$ affects the dissociation of heavy quarkonium. Our results show that $\lambda_{GB}$ reduces the peak height and increases the peak width of the spectral function, suggesting that $\lambda_{GB}$ enhances the dissociation of heavy quarkonium. We also discuss how the dissociation of heavy quarkonium varies with the ratio of shear viscosity to entropy density and find the dissociation will be easier in more perfect plasma. Additionally, we observe that the temperature decreases the peak height and widens the peak, thereby accelerating the dissociation.

Keywords: AdS/CFT correspondence, quarkonium dissociation, quark-gluon plasma
\end{abstract}

\maketitle

\section{Introduction}\label{sec:01_intro}

Experimental studies at the Relativistic Heavy Ion Collider (RHIC) and the Large Hadron Collider (LHC) offer unique opportunities to investigate quantum chromodynamics (QCD) matter under extreme conditions \cite{Arsene:2004fa,Adcox:2004mh,Back:2004je,Adams:2005dq}. The high energies produced in heavy ion collisions create a hot and dense environment, leading to the formation of a new state of matter called quark-gluon plasma (QGP) as expected due to the deconfinement phase transition. Heavy quarkonium ($J/\Psi$ and $\Upsilon(1S)$) plays a crucial role in probing the properties of QGP and provides valuable insights into the behavior of strongly coupled plasma. The suppression of heavy quarkoniums serves as evidence of their interaction with QGP and has garnered significant attention from researchers \cite{Matsui:1986dk,Satz:2005hx}. Thermal dissociation of heavy quarkonium refers to the process in which a particle peak in the spectral function disappears.

The results of lattice QCD suggest that $J/\Psi$ could exist as distinct resonances even up to temperatures around $1.6T_c$ ($T_c$ is the critical temperature), and then dissociate in the temperature range of $1.6T_c$ to $1.9T_c$ on anisotropic lattices \cite{Asakawa:2003re}. Further research indicates that on isotropic lattices, $J/\Psi$ gradually dissociates and disappears at around $3T_c$ according to lattice QCD \cite{Datta:2003ww}. Additionally, in a potential model, it has been found that the 1S state of charmonium dissociates at $1.2T_c$ and the excited states melt below $T_c$ \cite{Mocsy:2007jz}. Furthermore, lattice QCD predicts that the lower bound of the $\Upsilon(1S)$ state dissociates at around $2.3T_c$ \cite{Suzuki:2012ze}, and it suggests that the ground state of bottomonium could survive in a range from approximately $0.42T_c$ to $2.09T_c$ \cite{Aarts:2011sm}. The thermal behaviors of quarkonium states have been discussed using QCD sum rules \cite{Dominguez:2009mk, Dominguez:2013fca}.

AdS/CFT correspondence \cite{Maldacena:1997re,Witten:1998qj,Gubser:1998bc} could also provide valuable insights into the dissociation of heavy quarkonium. Studies on the dissociation of scalar glueballs and scalar mesons at finite temperature from a holographic perspective have been conducted \cite{Colangelo:2009ra,Miranda:2009uw}. Additionally, the influence of a magnetic field on the dissociation of heavy quarkonium has been explored in \cite{Dudal:2014jfa,Braga:2018zlu,Zhao:2021ogc}. The effects of temperature and chemical potential on the melting of heavy quarkonium have been analyzed in several studies \cite{Mamani:2013ssa,Braga:2016wkm,Braga:2017oqw,Braga:2017bml,Mamani:2022qnf}. Furthermore, discussions on the masses and decay constants of heavy quarkonium have been presented in \cite{Braga:2015jca,Braga:2015lck}. Studies have also been conducted on the quasinormal modes of heavy quarkonium within the framework of holographic QCD models \cite{Braga:2019yeh,Braga:2019xwl,Zhao:2023yry}. The analysis of the spectral functions of heavy quarkonium in a rotating background has been explored in \cite{Zhao:2023pne,Braga:2023fac,Zhu:2024uwu}. Moreover, the effects of anisotropy on the dissociation of heavy quarkonium have been discussed in \cite{Chang:2024ksq}. The spectra of light-flavor hadrons in the dynamical holographic QCD model have been examined in \cite{Chen:2022pgo}. For the dilepton decays of vector mesons, the authors of \cite{Sheng:2024kgg} calculate the relationship between production rates and the spectral function of mesons within the holographic model. The spectral function of the $J/\Psi$ meson in the soft wall model is discussed in \cite{Zhao:2024ipr}. The dissociation of scalar and vector mesons in two-flavor holographic QCD has been investigated in \cite{Ballon-Bayona:2024twa}. Additionally, the analysis of the spectral functions of fermions in instantonic plasma has been conducted in \cite{Li:2024apc}.

In string theory, it is widely recognized that there may be higher derivative corrections due to the complex interplay between string or quantum effects \cite{Douglas:2006es}. The corrections, such as the curvature squared corrections ($R^2$), are associated with the leading $1/N_c$ corrections when a D7-brane is present \cite{Aharony:1999rz,Aharony:2007dj,Buchel:2008vz}. The $R^2$ corrections can contribute to a more accurate description of interactions, including quantum gravitational effects or higher-dimensional operators. The $R^2$ term often represents higher derivative corrections or modifications to the action. The Gauss-Bonnet term naturally emerges when exploring higher-dimensional theories of gravity. In these frameworks, the action typically involves higher derivative corrections, and the Gauss-Bonnet term is one of the simplest corrections that can be made to the gravitational action. These higher derivative corrections are important for understanding various physical phenomena, including their impact on the ratio of shear viscosity to entropy density ($\eta/s$) in gravity theories with $R^2$ corrections \cite{Brigante:2007nu,Brigante:2008gz,Kats:2007mq}. Furthermore, higher derivative corrections have been studied in relation to the quark potential \cite{Noronha:2009ia,BitaghsirFadafan:2015yng}, drag force \cite{Fadafan:2008gb}, jet quenching parameter \cite{BitaghsirFadafan:2010rmb,Zhang:2015hkz}, and energy loss of light quarks \cite{Zhang:2023kzf}. The Gauss-Bonnet term contributes to the formation of black hole solutions, exhibiting various asymptotic behaviors and thermodynamic properties. In addition to the Graviton-Dilaton-Maxwell model, there are some researches on the QCD thermodynamics and QCD phase diagram through Einstien-Gauss-Bonnet gravity models in recent years \cite{Sajadi:2023zke,Sajadi:2023ckp,Hu:2024lxf}. The inclusion of the Gauss-Bonnet term offers new insights into the behavior of strongly-coupled gauge theories. Additionally, higher derivative corrections have been examined in the context of analytic structure of thermal correlators, transport coefficients \cite{Grozdanov:2016vgg,Grozdanov:2016fkt,Casalderrey-Solana:2018rle}, and other quantities \cite{Finazzo:2013rqy,Waeber:2015oka,Grozdanov:2016zjj,Li:2018lsl,Zhang:2018yzh,Buchel:2018eax,Waeber:2018bea,Folkestad:2019lam,Dutta:2022wbh}.

The authors of Ref. \cite{Braga:2016wkm} investigate the melting of heavy mesons in the context of AdS/QCD at finite temperature. However, the spectral functions of heavy quarkonium within Gauss-Bonnet gravity have not yet been explored. We want to expand upon the findings of Ref. \cite{Braga:2016wkm} by considering both finite temperature and finite $\lambda_{GB}$ cases. In this paper, we focus on studying $R^2$ corrections to the spectral functions of heavy quarkonium in the Gauss-Bonnet gravity and examining the influence of the Gauss-Bonnet parameter $\lambda_{GB}$ on the dissociation of heavy quarkonium. The results of Refs. \cite{Brigante:2007nu,Brigante:2008gz,Kats:2007mq} show that $\eta/s$ decreases as $\lambda_{GB}$ increases, indicating that the fluid behaves more like a perfect fluid when increasing $\lambda_{GB}$. Thus, studying the connection between $\eta/s$ and the dissociation of heavy quarkonium may be an interesting research.

The paper is organized as follows. In Sec.~\ref{sec:02}, we discuss the holographic model with higher derivative correction. In Sec.~\ref{sec:03}, we study the spectral function with higher derivative correction. In Sec.~\ref{sec:04}, we give the conclusion and discussion.

\section{Holographic model with higher derivative correction}\label{sec:02}

When the $R^2$ correction is present, the Gauss-Bonnet term is generally introduced as a correction to Einstein's general relativity. The $R^2$ corrections are associated with the leading $1/N_c$ corrections when a D7-brane is present \cite{Aharony:1999rz,Aharony:2007dj,Buchel:2008vz}. The effective action to leading order can be given as \cite{Brigante:2007nu,Brigante:2008gz}
\begin{equation}\label{eq:x1}
\begin{split}
I=& \frac{1}{16\pi G_{5}} \int d^{5}x\sqrt{-g} \bigg[R+\frac{12}{L^2}+ L^2 (m_{1}R^2 +m_{2}R_{\mu\nu}R^{\mu\nu} +m_{3}R_{\mu\nu\rho\sigma}R^{\mu\nu\rho\sigma})\bigg],
 \end{split}
\end{equation}
where $R$, $R_{\mu\nu}$ and $R_{\mu\nu\rho\sigma}$ denote Ricci scalar, Ricci tensor and Riemann tensor respectively. $G_{5}$ represents Newton constant. In higher-dimensional theories, adding the Gauss-Bonnet term is important for ensuring consistency and stability in the theory. This addition helps avoid certain types of singularities in the equations of motion and guarantees that the theory remains well-defined. $L$ denotes AdS radius at leading order in $m_{i}$ , where $m_{i}\sim \frac{\alpha'}{L^2}\ll 1$. It should be mentioned that only $m_3$ is unambiguous, while $m_1$ and $m_2$ could be arbitrarily altered from the field redefinition \cite{Brigante:2007nu,Brigante:2008gz}. In order to avoid this issue, one can consider the Gauss-Bonnet gravity, where $m_{i}$ are fixed by parameter $\lambda_{GB}$. Gauss-Bonnet gravity is one special case of the action (1), and the action with $\lambda_{GB}$ term in Gauss-Bonnet gravity can be given as
\begin{equation}\label{eq:x2}
\begin{split}
I=&\frac{1}{16\pi G_{5}} \int d^{5}x\sqrt{-g} \bigg[R+\frac{12}{L^2}+\frac{\lambda_{GB}}{2}L^2 (R^2 -4R_{\mu\nu}R^{\mu\nu} +R_{\mu\nu\rho\sigma}R^{\mu\nu\rho\sigma})\bigg],
 \end{split}
\end{equation}
where $\lambda_{GB}$ is valid in $-\frac{7}{36}< \lambda_{GB}\leq \frac{9}{100}$ range. The lower bound of the range is from the requirement that the boundary energy density should be positive-definite \cite{Hofman:2008ar}.

The metric of Gauss-Bonnet gravity is \cite{Cai:2001dz}
\begin{equation}
\label{eqa1}
 ds^2= \frac{L^2}{z^2}\bigg(-n^2f(z)dt^2 + d\overrightarrow{x}^2 + \frac{1}{f(z)}dz^2 \bigg),
\end{equation}
with
\begin{equation}
\label{eqa2}
 f(z)= \frac{1}{2 \lambda_{GB}} \bigg(1-\sqrt{1-4\lambda_{GB}(1-\frac{z^4}{z^4_h})} \bigg),
\end{equation}
and
\begin{equation}
\label{eqa3}
 n^2 =\frac{1}{2}(1+\sqrt{1-4\lambda_{GB}}),
\end{equation}
where $\overrightarrow{x}$ denotes the boundary coordinates $x_1$, $x_2$ and $x_3$. $z$ is the coordinate of fifth dimension. $z_h$ denotes the horizon. The temperature is \cite{Cai:2001dz}
\begin{equation}
\label{eqa4}
 T =\frac{n}{\pi z_h}.
\end{equation}

\section{Spectral function with higher derivative correction}\label{sec:03}
In this section, we will examine the spectral function of charmonium and bottomonium in Gauss-Bonnet gravity. The formulas for the spectral function can be derived by referring to the work in Ref. \cite{Braga:2018zlu}. We can represent heavy quarkonium using the vector field $V_m=(V_\mu, V_z)$, which is dual to the gauge theory current $J^\mu=\overline{\Psi}\gamma^\mu\Psi$. The gravitational action is defined as follows \cite{Braga:2018zlu}:
\begin{equation}\label{eq:x3}
I= \int d^{4}x dz \sqrt{-g} e^{-\phi(z)}\bigg[-\frac{1}{4g^2_5 }F_{mn}F^{mn} \bigg],
\end{equation}
where $F_{mn}=\partial_m V_n -\partial_n V_m$. We assume the value of $g^2_5$ to be one for simplicity in this work. The dilaton field $\phi(z)$ can choose the following form \cite{Braga:2018zlu}
\begin{equation}\label{eq:x4}
\phi(z)= w^2 z^2+Mz+\tanh(\frac{1}{Mz}-\frac{w}{\sqrt{\Gamma}}),
\end{equation}
where $w$ and $\Gamma$ denote the quark mass and string tension respectively. $M$ represents the mass scale which is related to the non-hadronic decay. These parameters can be fixed by fitting the spectrum of masses \cite{Braga:2018zlu}. The values of these parameters for charmonium and bottomonium are
\begin{equation}
\label{eqav5}
\begin{split}
&  w_c=1.2GeV, \sqrt{\Gamma_c}=0.55GeV, M_c = 2.2GeV;\\
 & w_b=2.45GeV, \sqrt{\Gamma_b}=1.55GeV, M_b = 6.2GeV.
 \end{split}
\end{equation}

The spectral function of heavy mesons can be obtained from membrane paradigm \cite{Iqbal:2008by}. We rewrite the background form (Eq.(\ref{eqa1})) of Gauss-Bonnet gravity
\begin{equation}
\label{eqa6}
 ds^2= -g_{tt}dt^2 +g_{xx_{1}} dx_{1}^{2}+ g_{xx_{2}} dx_{2}^{2}+g_{xx_{3}} dx_{3}^{2}+g_{zz}dz^2.
\end{equation}

From Eq.(\ref{eq:x3}), one can get the equation of motion
\begin{equation}
\label{eqa8}
 \partial^m \bigg(\frac{\sqrt{-g}}{h(z)}F_{mn}\bigg)=0,
\end{equation}
where $h(z)=e^{\phi(z)}$.

The conjugate momentum of the gauge field in the case of z-foliation is
\begin{equation}
\label{eqa9}
 j^\mu= -Q F^{z\mu}.
\end{equation}

The background of Gauss-Bonnet gravity in $\overrightarrow{x}$-direction is isotropic. Here we can consider that the equations of motion has longitudinal fluctuations along ($t, x_3$) and transverse fluctuations along ($x_1, x_2$). The longitudinal components of Eq.(\ref{eqa8}) are
\begin{equation}
\label{eqav8}
\begin{split}
&  -\partial_z j^t -\frac{\sqrt{-g}}{h(z)}g^{tt}g^{xx_3}\partial_{x_3}F_{x_3 t}=0,\\
 & -\partial_z j^{x_3} +\frac{\sqrt{-g}}{h(z)}g^{tt}g^{xx_3}\partial_{t}F_{x_3 t}=0.\\
 & \partial_{x_3} j^{x_3}+\partial_t j^t =0,
 \end{split}
\end{equation}

Applying the Bianchi identity, one can get
\begin{equation}
\label{eqa9}
 \partial_z F_{x_3 t}-\frac{h(z)}{\sqrt{-g}}g_{zz}g_{xx_3}\partial_t j^z -\frac{h(z)}{\sqrt{-g}} g_{tt}g_{xx_3}\partial_{x_3}j^t=0.
\end{equation}

The conductivity of longitudinal channel and its derivative are
\begin{equation}
\label{eqav9}
\begin{split}
&  \overline{\sigma}_L (\omega,\overrightarrow{p},z)=\frac{j^{x_3}(\omega,\overrightarrow{p},z)}{F_{x_3 t}(\omega,\overrightarrow{p},z)},\\
 & \partial_z \overline{\sigma}_L=-i\omega\sqrt{\frac{g_{zz}}{g_{tt}}}\bigg[\Sigma(z)-\frac{\overline{\sigma}^2_L}{\Sigma(z)}(1-\frac{p^2_3}{\omega^2}\frac{g^{xx_3}}{g^{tt}}) \bigg],
 \end{split}
\end{equation}
where the momentum $p=(\omega,0,0,p_3)$ and $\Sigma(z)=\frac{1}{h(z)}\sqrt{\frac{-g}{g_{zz}g_{tt}}}g^{xx_3}$.

In the same way, one can get the result of transverse channel
\begin{equation}
\label{eqa10}
 \partial_z \overline{\sigma}_T=i\omega \sqrt{\frac{g_{zz}}{g_{tt}}} \bigg[\frac{\overline{\sigma}^2_T}{\Sigma(z)}- \Sigma(z)(1-\frac{p^2_3}{\omega^2}\frac{g^{xx_3}}{g^{tt}})  \bigg].
\end{equation}

In zero momentum ($p^2_3 =0$) case, one can find $ \overline{\sigma}= \overline{\sigma}_T = \overline{\sigma}_L$ and the form is
\begin{equation}
\label{eqa11}
 \partial_z \overline{\sigma}=i\omega \sqrt{\frac{g_{zz}}{g_{tt}}} \bigg[\frac{\overline{\sigma}^2}{\Sigma(z)}- \Sigma(z)  \bigg].
\end{equation}

From the Kubo formula, one can find AC conductivity $\sigma$ is relevant with Retarded Green function
\begin{equation}
\label{eqa12}
 \sigma(\omega)=-\frac{G_R (\omega)}{i \omega}\equiv \overline{\sigma}(\omega,z=0).
\end{equation}

Then one can get the spectral function
\begin{equation}
\label{eqa131}
 \rho(\omega)\equiv -Im G_R (\omega)=\omega Re \overline{\sigma}(\omega,0).
\end{equation}

\begin{figure}[H]
    \centering
      \setlength{\abovecaptionskip}{0.1cm}
    \includegraphics[width=14cm]{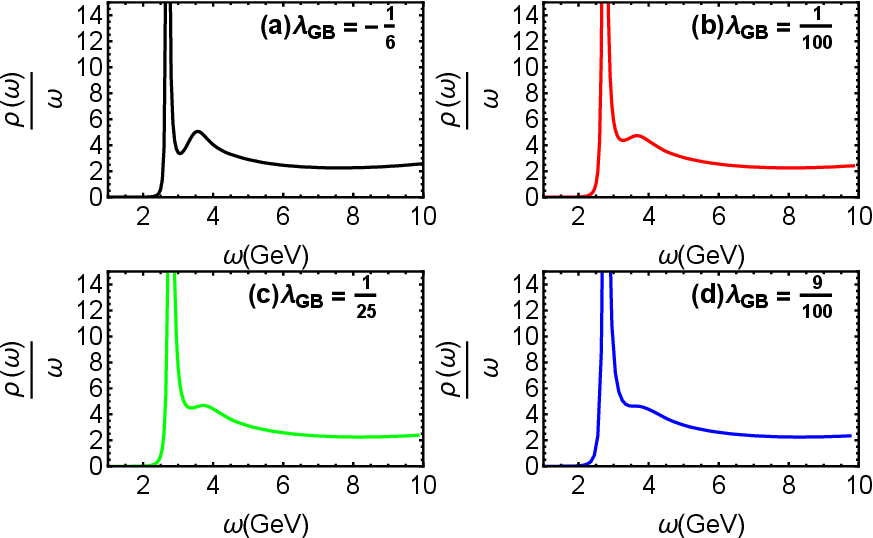}
    \caption{\label{fig1} Spectral functions of charmonium at $T=0.15$ GeV for different values of $\lambda_{GB}$.}
\end{figure}

Using the parameters of Eq.(\ref{eqav5}), we can calculate the spectral function of charmonium and bottomonium with numerical method. In calculations, we set AdS radius $L$ to be 1.

In Fig.~\ref{fig1}, we present the spectral functions for charmonium at $T=0.15$ GeV for various values of $\lambda_{GB}$. The quasiparticle state is characterized by the bell-shaped profile of the spectral function. The first and second peaks correspond to the 1S state ($J/\Psi$) and the 2S state, respectively. From the results shown in Fig.~\ref{fig1}, it is evident that increasing $\lambda_{GB}$ leads to a reduction of the 2S state.

\begin{figure}[H]
    \centering
      \setlength{\abovecaptionskip}{0.1cm}
    \includegraphics[width=9cm]{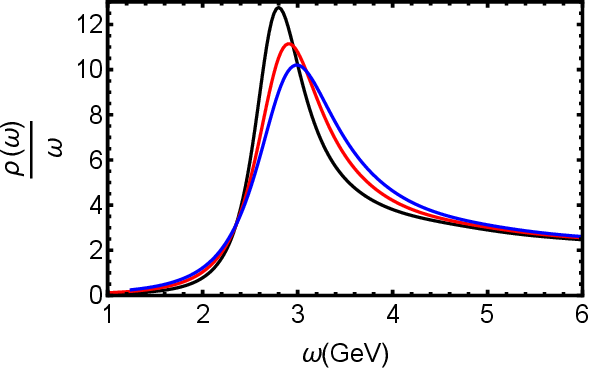}
    \caption{\label{fig2} Spectral functions of charmonium at $T=0.3$ GeV for different values of $\lambda_{GB}$. The black, red and blue line denote $\lambda_{GB} = -\frac{1}{6},\ \frac{1}{100}$, and $\frac{9}{100}$, respectively.}
 \end{figure}

In Fig.~\ref{fig2}, we plot the spectral functions for charmonium at $T=0.3$ GeV for different values of $\lambda_{GB}$. The peak width is inversely proportional to decay rate, suggesting the stability. It is obvious that increasing $\lambda_{GB}$ causes the spectral function peak to decrease in height and broaden in width, indicating the promotion of the dissociation effect of $J/\Psi$.

\begin{figure}[H]
    \centering
      \setlength{\abovecaptionskip}{0.1cm}
    \includegraphics[width=14cm]{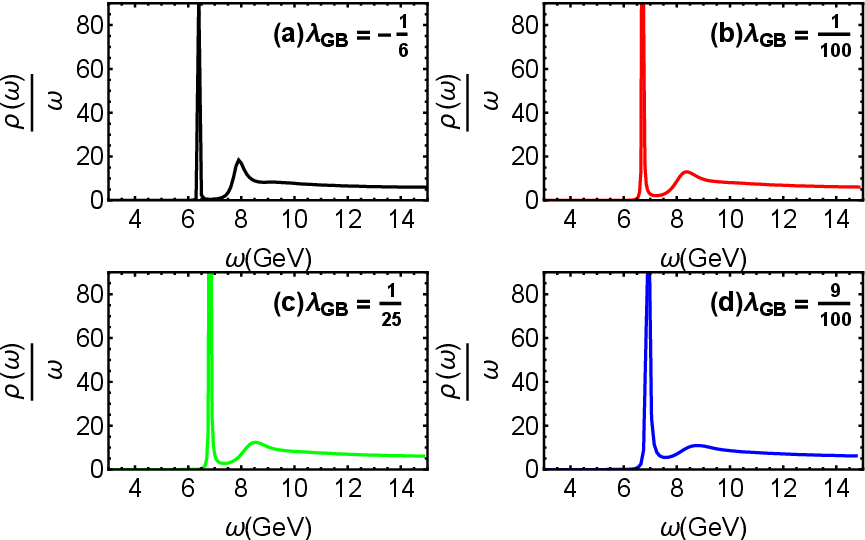}
    \caption{\label{fig3} Spectral functions of bottomonium at $T=0.2$ GeV for different values of $\lambda_{GB}$. }
\end{figure}

In Fig.~\ref{fig3}, we present the spectral functions for bottomonium at $T=0.2$ GeV with different values of $\lambda_{GB}$. The first and second peaks represent the 1S state ($\Upsilon(1S)$) and the 2S state, respectively. The results shown in Fig.~\ref{fig3} indicate that the 2S state gradually diminishes as $\lambda_{GB}$ increases.

\begin{figure}[H]
    \centering
      \setlength{\abovecaptionskip}{0.1cm}
    \includegraphics[width=9cm]{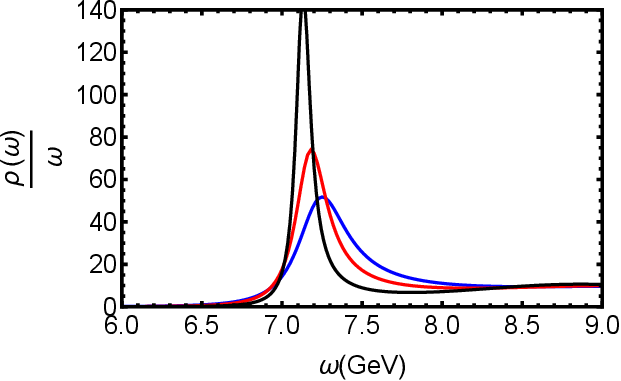}
    \caption{\label{fig4} Spectral functions of bottomonium at $T=0.3$ GeV for different values of $\lambda_{GB}$. The black, red and blue line denote $\lambda_{GB} = -\frac{1}{6},\ \frac{1}{100}$, and $\frac{9}{100}$, respectively.}
\end{figure}

In Fig.~\ref{fig4}, we show the spectral functions for bottomonium at $T=0.3$ GeV for various $\lambda_{GB}$ values. The peak corresponds to the 1S state ($\Upsilon(1S)$). We observed that the peak height decreases as $\lambda_{GB}$ increases, while the peak width increases with increasing $\lambda_{GB}$. It indicates that $\lambda_{GB}$ enhances the dissociation effect of $\Upsilon(1S)$.

From the results of Figs.~\ref{fig1} - ~\ref{fig4}, we can conclude that $\lambda_{GB}$ promotes the dissociation of heavy quarkonium in Gauss-Bonnet gravity. In the works of Refs. \cite{Brigante:2007nu,Brigante:2008gz,Kats:2007mq}, the authors investigate how the parameter $\lambda_{GB}$ affects the ratio of shear viscosity to entropy density ($\eta/s$) in Gauss-Bonnet gravity, $\frac{\eta}{s}=\frac{1}{4\pi}(1-4\lambda_{GB})$. They find that $\eta/s$ decreases as $\lambda_{GB}$ increases, indicating that the fluid behaves more like a perfect fluid when increasing $\lambda_{GB}$. Consequently, we can conclude that the dissociation of heavy quarkonium will be easier in more perfect plasma. Moreover, $\eta/s$ decreases $38.56\% $ when increasing $\lambda_{GB}$ from $-\frac{1}{6}$ to $\frac{9}{100}$. The peak of $J/\Psi$ decreases $24\% $ and the $\Upsilon(1S)$ decreases $64.29\% $ when increasing $\lambda_{GB}$ from $-\frac{1}{6}$ to $\frac{9}{100}$.

In Ref. \cite{Zhao:2023yry}, the authors calculate the configuration entropy of bottomonium in Gauss-Bonnet gravity and find that $\lambda_{GB}$ enhances the dissociation effect. In Ref. \cite{Zhang:2018yzh}, the authors study the impact of $\lambda_{GB}$ on the entropic force of the heavy quarkonium. The entropic force can be seen as one mechanism that melts the heavy quarkonium. It is found that $\lambda_{GB}$ increases the entropic force and enhances the dissociation of heavy quarkonium. The qualitative results of our work are consistent with the results of Refs. \cite{Zhao:2023yry,Zhang:2018yzh}.

\begin{figure}[H]
    \centering
      \setlength{\abovecaptionskip}{0.1cm}
    \includegraphics[width=14cm]{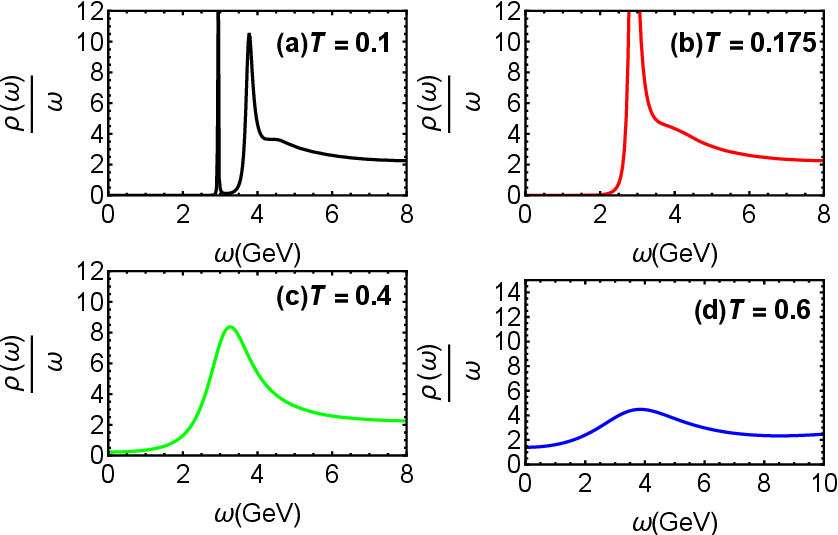}
    \caption{\label{fig5} Spectral functions of charmonium at $\lambda_{GB} = \frac{1}{100}$ for different values of $T$. $T$ is in units $GeV$.}
\end{figure}

\begin{figure}[H]
    \centering
      \setlength{\abovecaptionskip}{0.1cm}
    \includegraphics[width=14cm]{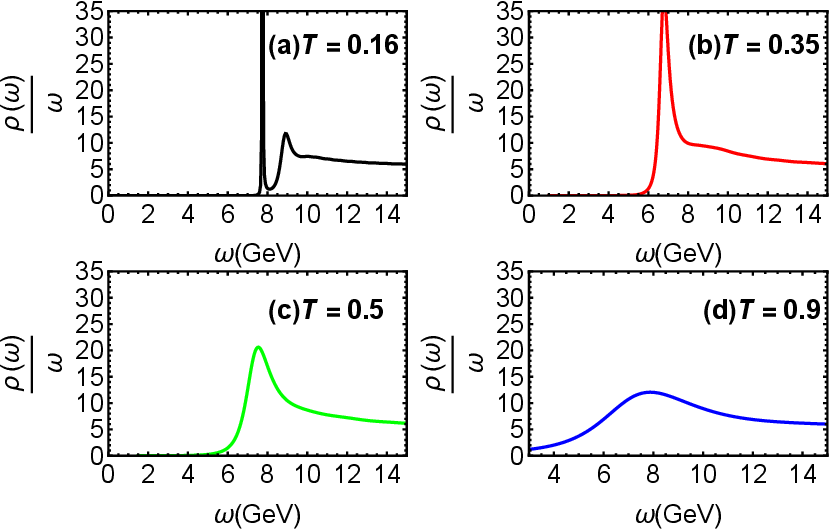}
    \caption{\label{fig6} Spectral functions of bottomonium at $\lambda_{GB} = \frac{1}{100}$ for different values of $T$. $T$ is in units $GeV$.}
\end{figure}

In the work of \cite{Sachan:2013zza}, the authors examine the confinement/deconfinement transition within the soft wall model, incorporating Gauss-Bonnet corrections. They determine that the phase transition temperature $T_c \sim 0.19GeV$ at zero chemical potential and zero Gauss-Bonnet term. The authors also note that the effect of the Gauss-Bonnet term on the phase transition temperature is minimal. Therefore, for the purposes of this study, we will adopt the phase transition temperature $T_c \sim 0.19GeV$ at $\lambda_{GB} = \frac{1}{100}$ for simplicity.

In Fig.~\ref{fig5}, we analyze the spectral functions of charmonium at $\lambda_{GB} = \frac{1}{100}$ for different temperatures. From Fig.~\ref{fig5} (a), we observe that the 2S state and $J/\Psi$ coexist at low temperature. As $\lambda_{GB}$ increases, the 2S state gradually disappears, with total melting occurring at $T= 0.175GeV$ ($T=0.92T_c$) as shown in Fig.~\ref{fig5} (b). This finding is consistent with results from the potential model \cite{Mocsy:2007jz}. Additionally, the $J/\Psi$ also disappears as temperature increases, with complete melting occurring at $T= 0.6GeV$ ($T=3.16T_c$), as illustrated in Figs.~\ref{fig5} (c) and (d). This result aligns closely with findings from lattice QCD \cite{Datta:2003ww}.

In Fig.~\ref{fig6}, we examine the spectral functions of bottomonium at \(\lambda_{GB} = \frac{1}{100}\) for various temperatures. From Fig.~\ref{fig6} (a), we observe that the 2S state and \(\Upsilon(1S)\) coexist at low temperature. As \(\lambda_{GB}\) increases, the 2S state gradually diminishes, with total melting happening at $T= 0.35GeV$ ($T=1.84T_c$), as shown in Fig.~\ref{fig6} (b). This indicates that the $\Upsilon(1S)$ also begins to melt at $T=1.84T_c$, which closely aligns with the results from lattice QCD \cite{Suzuki:2012ze}. Furthermore, the $\Upsilon(1S)$ continues to disappear as the temperature rises, with complete melting occurring at $T= 0.9GeV$ ($T=4.74T_c$), as shown in Figs.~\ref{fig6} (c) and (d). From the results of Fig.~\ref{fig5} and Fig.~\ref{fig6}, one can summarize that the temperature promotes the dissociation effect of heavy quarkonium.

\section{Conclusion and discussion}\label{sec:04}

In this paper, we investigate the $R^2$ corrections to the spectral functions in Gauss-Bonnet gravity, specifically focusing on the effect of the Gauss-Bonnet parameter $\lambda_{GB}$ on the dissociation of heavy quarkonium.

It is clear that increasing $\lambda_{GB}$ causes the peak of the spectral functions to decrease in height and broaden in width. These findings indicate that the dissociation effect of heavy quarkonium is promoted. Thus, we can conclude that $\lambda_{GB}$ promotes the dissociation of heavy quarkonium in Gauss-Bonnet gravity. Moreover, it is observed that as the temperature increases, the peak height decreases, and the peak width increases. This suggests that the temperature promotes the dissociation effect of heavy quarkonium. We also discuss how the dissociation of heavy quarkonium varies with the ratio of shear viscosity to entropy density and find the dissociation will be easier in more perfect plasma.

We expect this work can provide valuable insights in the behavior of strongly coupled plasma. Finally, it is also interesting to study the effect of $R^4$ corrections on the spectral functions. We hope to report this work in the future.

\section*{Acknowledgments}

Rui Zhou is supported by the Science and Technology Development Plan Project of Henan Province No. 242102230085. Rui Zhou is supported by the National Natural Science Foundation of China under Grant No. 12404470. Rui Zhou is also supported by the High Level Talents Research and Startup Foundation Projects for Doctors of Zhoukou Normal University No. ZKNUC2023017. Manman Sun is supported by the National Natural Science Foundation of China under Grant No.12305076. Zhou-Run Zhu is supported by the High Level Talents Research and Startup Foundation Projects for Doctors of Zhoukou Normal University No. ZKNUC2023018.


\end{document}